\documentclass[11pt,a4paper]{article}
\pdfoutput=1
\usepackage{jheppub}

\usepackage{graphicx,bm}
\usepackage{amsmath,amssymb,mathrsfs,verbatim,subfigure}

\title{Holographic RG flow of the shear viscosity to entropy density ratio in strongly coupled anisotropic plasma}

\author{Kiminad A. Mamo}

\affiliation{Department of Physics, University of Illinois,
Chicago, IL 60607-7059, USA}

\emailAdd{kabebe2@uic.edu}

\abstract{

  We study holographic RG flow of the shear viscosity tensor of anisotropic, strongly coupled $\mathcal{N}=4$ super-Yang-Mills plasma by using its type IIB supergravity dual in anisotropic bulk spacetime. We find that the shear viscosity tensor has three independent components in the anisotropic bulk spacetime away from the boundary, and one of the components has a non-trivial RG flow while the other two have a trivial one. 
  For the component of the shear viscosity tensor with non-trivial RG flow, we derive its RG flow equation, and solve the equation analytically to second order in the anisotropy parameter $a$. We derive the RG equation using the equation of motion, holographic Wilsonian RG method, and Kubo's formula. All methods give the same result. Solving the equation, we find that the ratio of the component of the shear viscosity tensor to entropy density $\frac{\eta}{s}$ flows from above $\frac{1}{4\pi}$ at the horizon (IR) to below $\frac{1}{4\pi}$ at the boundary (UV) where it violates the holographic shear viscosity (Kovtun-Son-Starinets) bound and where it agrees with the other longitudinal component.
}


\arxivnumber{}

\begin{document}

\maketitle

\section{Introduction}

AdS/CFT correspondence \cite{Maldacena:1997re, Gubser:1998bc, Witten:1998qj} is a useful theoretical tool in order to calculate field theory predictions at strong coupling directly from their weakly coupled classical gravity duals or low energy string theory at large-$N$. The string theory or gravity dual is formulated in a bulk spacetime which asymptotes to Anti-de Sitter (AdS) spacetime. The bulk spacetime comes with an extra radial dimension which can be interpreted as the energy scale of the field theory \cite{Maldacena:1997re, Susskind:1998dq}. The on-shell action of the gravity theory contains some terms which diverge at the boundary, and this corresponds to UV divergences of the field theory. And, just like we would in field theory, we should introduce a renormalization scheme in the gravity theory side in order to eliminate these divergences. This is done, in the gravity side, by using the holographic renormalization procedure \cite{de Haro:2000xn}. By implementing this procedure, one can, for example, calculate the renormalized two-point correlation functions of the energy-momentum tensor $\langle T^{b}\,_{a}T^{b}\,_{a}\rangle$ which can in turn be used to calculate the transport coefficients, like the shear viscosity tensor $\eta^{b}\,_{a}\,^{b}\,_{a}$, of a fluid using Kubo's formula \cite{Son:2002sd}. (Note: Throughout this paper the indices $\{a,b,c\}$ run over all the spatial coordinates $x$, $y$, and $z$ while $\{i,j\}$ run over the spatial coordinates $x$, and $y$ only.) If there's is no conformal anomaly, the shear viscosity tensor calculated in this fashion, will be independent of the radial direction, and hence energy scale.

If there is a conformal anomaly, however, the holographically renormalized two-point function runs with energy scale according to Callan-Symanzik renormalization group (RG) equation \cite{de Haro:2000xn}. Therefore, in this case, one expects the shear viscosity tensor also to run, and its value at the boundary (UV) to be different from the one at the horizon (IR). But, if there is no conformal anomaly, the Callan-Symanzik RG flow equation of the two-point function will be trivial. Hence, the shear viscosity tensor won't run, and its value at the boundary (UV) will be the same as the one at the horizon (IR). This has been checked, for example, for isotropic bulk spacetime, where there is no conformal anomaly, by independent calculations of the shear viscosity tensor at the horizon (IR), which goes by the name of 'membrane paradigm' \cite{Kovtun:2003wp,Iqbal:2008by}, and earlier works in 1970s \cite{Damour:1979de}, which resulted in a value exactly the same as the one at the boundary (UV) \cite{Son:2002sd} which was calculated by implementing the holographic renormalization procedure. This was later confirmed by directly deriving the RG flow equation for the shear viscosity tensor \cite{Iqbal:2008by}, by using the equation of motion for the gravitational fluctuations in isotropic bulk spacetime, which turned out to be trivial, therefore, the shear viscosity tensor took the same value at any energy scale.

Since, the strongly coupled quark gluon plasma created in the heavy ion collision \cite{Adams:2005dq} is anisotropic \cite{Florkowski:2008ag}, it's important to study the anisotropic version of $\mathcal{N}=4$ $SU(N_{c})$ super-Yang-Mills plasma by using its type IIB supergravity dual \cite{Azeyanagi:2009pr,Mateos:2011tv}. For example, the trace of the energy-momentum tensor of the anisotropic $\mathcal{N}=4$ plasma has been calculated in \cite{Mateos:2011tv}, by using its gravity dual, and it turned out to be proportional to the anisotropy parameter $a$, more precisely, $\langle T^{b}\,_{b}\rangle=\frac{N^{2}_{c}a^{4}}{48\pi^{4}}$. This shows that there is conformal anomaly in the anisotropic $\mathcal{N}=4$ plasma due to the anisotropy. Hence, the Callan-Symanzik RG flow equation for the two-point function, consequently, the RG flow of some components of the shear viscosity tensor $\eta^{b}\,_{a}\,^{b}\,_{a}$ must be non-trivial. In fact, in this paper, we show that an independent component of the shear viscosity tensor, $\eta^{i}\,_{z}\,^{i}\,_{z}$, has a non-trivial RG flow while the other independent components of the shear viscosity tensor, $\eta^{j}\,_{i}\,^{j}\,_{i}$ and $\eta^{z}\,_{i}\,^{z}\,_{i}$, have a trivial RG flow.

The fact that we have three independent components, $\eta^{j}\,_{i}\,^{j}\,_{i}$, $\eta^{z}\,_{i}\,^{z}\,_{i}$, and $\eta^{i}\,_{z}\,^{i}\,_{z}$, in the anisotropic bulk spacetime away from the boundary is the result of the antisymmetry of one index up and one index down energy-momentum tensor operator there, i.e.,
$\gamma^{ii}(\epsilon\neq0)T_{iz}=T^{i}\,_{z}\neq T^{z}\,_{i}=\gamma^{zz}(\epsilon\neq 0)T_{zi}$. (Note that: the boundary is located at $u=\epsilon=0$, and the indices of the energy-momentum tensor operator at the hypersurface away from the boundary $\epsilon\neq 0$ are raised using the induced metric there, i.e., $\gamma^{ii}(\epsilon\neq 0)\neq \gamma^{zz}(\epsilon\neq 0)$). Therefore, at the boundary, where the energy-momentum tensor operator is symmetric, since $\gamma^{ii}(\epsilon=0)=\gamma^{zz}(\epsilon=0)$, we expect to have only two independent components of the shear viscosity tensor: $\eta^{j}\,_{i}\,^{j}\,_{i}$ and $\eta^{z}\,_{i}\,^{z}\,_{i}$. (Nonsymmetric one index up and one index down energy-momentum tensor operator was also noted in a different context in reference \cite{Adams:2008wt}, sect. 4.2.)

In this paper, we derive the non-trivial holographic RG flow equation of the shear viscosity $\eta^{i}\,_{z}\,^{i}\,_{z}$ of the anisotropic $\mathcal{N}=4$ plasma, using the equation of motion for the shear mode gravitational fluctuations \cite{Iqbal:2008by}, the holographic Wilsonian RG method \cite{Faulkner:2010jy,Heemskerk:2010hk,Sin:2011yh,Grozdanov:2011aa}, and Kubo's formula, and give analytical solutions up to first order in the anisotropy parameter $a$. From the solution of the RG flow equation, we find that at the boundary (UV) $\eta^{i}\,_{z}\,^{i}\,_{z}(\epsilon=0)$ is equivalent to $\eta^{z}\,_{i}\,^{z}\,_{i}(\epsilon=0)$ which is consistent with the fact that the one index up and one index down energy-momentum tensor operator at the boundary is symmetric, and the shear viscosity tensor has only two independent components: $\eta^{j}\,_{i}\,^{j}\,_{i}$ and $\eta^{z}\,_{i}\,^{z}\,_{i}$ \cite{Erdmenger:2010xm}.

Non-trivial holographic RG flow for some components of the shear viscosity tensor has also been reported in other anisotropic systems with gravity dual. Recent work \cite{Oh:2012zu}, in anisotropic superfluids \cite{Basu:2009vv}, has shown that the holographic RG flow of some components of the shear viscosity tensor is non-trivial.

Finite $N$ and higher derivative corrections to the weakly coupled two derivative gravity dual models in isotropic bulk spacetime result in a temperature dependent corrections to the shear viscosity tensor \cite{Kovtun:2011np, Myers:2009ij}. However, the holographic RG flow of the shear viscosity tensor is still trivial in this models. See \cite{Cremonini:2011iq} for recent review.

The outline of this paper is as follows: In section \ref{sec:iso}, we review reference \cite{Iqbal:2008by}'s derivation of the holographic RG flow equation for the shear viscosity tensor $\eta^{b}\,_{a}\,^{b}\,_{a}$ in a general isotropic bulk spacetime. We find that the RG flow equations of all components of $\eta^{b}\,_{a}\,^{b}\,_{a}$ are trivial in the hydrodynamic limit, and their initial value is determined by requiring regularity at the horizon. We also show that the shear viscosity tensor to entropy density ratio $\frac{\eta^{b}\,_{a}\,^{b}\,_{a}}{s}$ is given in terms of a ratio of the metric components, i.e., $\frac{\eta^{b}\,_{a}\,^{b}\,_{a}}{s}=\frac{1}{4\pi}\frac{g_{aa}}{g_{bb}}$, which can also be derived using the membrane paradigm approach. However, since $g_{aa}=g_{bb}$ in isotropic spacetime, $\frac{\eta^{b}\,_{a}\,^{b}\,_{a}}{s}$ always takes the universal value $\frac{1}{4\pi}$.

In section \ref{sec:aniso}, we derive the holographic RG flow equation for the shear viscosity tensor $\eta^{b}\,_{a}\,^{b}\,_{b}$ in anisotropic bulk spacetime which is a solution of type IIB supergravity, and show that the RG flow equations for its component $\eta^{i}\,_{z}\,^{i}\,_{z}$ is no more trivial in the hydrodynamic limit, even though, for $\eta^{b}\,_{i}\,^{b}\,_{i}$, it is still trivial. We also show that similar to the isotropic spacetime, in the anisotropic spacetime, the initial value of $\frac{\eta^{b}\,_{a}\,^{b}\,_{a}}{s}$ or its value at the horizon is determined by requiring regularity at the horizon $\epsilon=u_{h}$, and it is given by the ratio of the metric components, i.e., $\frac{\eta^{b}\,_{a}\,^{b}\,_{a}(\epsilon=u_{h})}{s}=\frac{1}{4\pi}\frac{g_{aa}(u_{h})}{g_{bb}(u_{h})}$, which can also be found by applying the membrane paradigm approach. Moreover, we observe that at the boundary $\frac{\eta^{z}\,_{i}\,^{z}\,_{i}(\epsilon=0)}{s}$ is equivalent to $\frac{\eta^{i}\,_{z}\,^{i}\,_{z}(\epsilon=0)}{s}$, even though, they are not equivalent in the anisotropic bulk spacetime away from the boundary, i.e., at $\epsilon\neq 0$.

In section \ref{sec:soln}, we solve the the non-trivial holographic RG flow equation of $\eta^{i}\,_{z}\,^{i}\,_{z}$ and find that $\frac{\eta^{i}\,_{z}\,^{i}\,_{z}}{s}$ flows from above $\frac{1}{4\pi}$ at the horizon (IR), which we find by imposing the regularity condition at the horizon, to below $\frac{1}{4\pi}$ at the boundary (UV) where it violates the holographic shear viscosity bound (Kovtun-Son-Starinets bound \cite{Kovtun:2004de}).

In Appendix \ref{sec:wilson}, and Appendix \ref{sec:kubo} we re-derive the holographic RG flow equation of $\eta^{i}\,_{z}\,^{i}\,_{z}$, already derived in section \ref{sec:aniso}, this time using the holographic Wilsonian RG method, and Kubo's formula, respectively.

\section{\label{sec:iso}Holographic RG flow in isotropic bulk spacetime}
In this section, we review the holographic RG flow of the shear viscosity tensor in a general isotropic bulk spacetime following closely reference \cite{Iqbal:2008by}.
\subsection{Effective action for the gravitational shear mode fluctuations in isotropic bulk spacetime}
As shown in \cite{Kovtun:2003wp}, later in \cite{Iqbal:2008by}, and more recently in \cite{Matsuo:2011fk} the relevant equations for gravitational shear mode fluctuations can be mapped onto an electromagnetic problem. Consider a metric perturbation of the form
 \begin{equation}
g_{aN}(r)\rightarrow g_{aN}(r)+g_{aa}h^{a}\,_{N}(x_{M}\neq a)
\end{equation}
in isotropic bulk spacetime
\begin{equation}
ds^{2}=g_{MN}dx^{M}dx^{N}=g_{tt}dt^{2}+g_{aa}dx^{a}dx^{a}+g_{uu}du^{2}.
\end{equation}
Indices: $\{L, M, N,\}$ run over the full 5-dimensional bulk; $\{a,b,c\}$ run over all spatial coordinates $x$, $y$, and $z$. And, throughout this paper the Einstein summation convention will apply only for indices $\{L, M, N,\}$ but not for $\{a,b,c\}$. Comparing this to the standard problem of Kaluza-Klein dimensional reduction along the $a$ spatial direction, setting $A_{N}^{a}\equiv h^{a}\,_{N}$, and using the gauge $h_{NN}=h_{uN}=0$, the Einstein-Hilbert action
\begin{equation}
S_{\rm bulk}=\frac{1}{2\kappa^2}\int_{\mathcal{M}}\!\!\sqrt{-g}\,R,
\end{equation}
after expanding it to second order in the gravitational shear mode fluctuations $h^{a}\,_{N}$, with gravitational coupling $\frac{1}{2\kappa^{2}}=\frac{1}{16\pi G}$, can be mapped onto Maxwell's action for the gauge fields $A_{N}^{a}$, with an effective gauge coupling $\frac{1}{g_{eff_a}^{2}}=\frac{1}{2\kappa^{2}}g_{aa}=\frac{1}{2\kappa^{2}}g_{xx}=\frac{1}{2\kappa^{2}}g_{yy}=\frac{1}{2\kappa^{2}}g_{zz}$ \cite{Kovtun:2003wp,Iqbal:2008by}
\begin{eqnarray}\label{isoaction}
  S_{eff} &=& -\frac{1}{4}\int d^{5}x\mathcal{N}^{MN}_{a}F^{a}_{MN}F^{a}_{MN},
\end{eqnarray}
where
\begin{eqnarray}
F^{a}_{MN} &=& \partial_{M}A^{a}_{N}-\partial_{N}A^{a}_{M}, \\
 A^{a}_{N} &=& h^{a}\,_{N}=g^{aa}h_{aN}(t,u,c\neq a), \\
 \mathcal{N}^{MN}_{a}(u) &=& \frac{1}{2\kappa^{2}}g_{aa}\sqrt{-g}g^{MM}g^{NN}.
 \end{eqnarray}
The effective action for $A^{a}_{N}$ with the effective gauge coupling $g_{eff_a}$ can be further mapped on to an action for scalar fields
$\psi^{a}_{b}\equiv A_{b}^{a}$
\begin{equation}\label{isosaction}
S_{eff}=-\frac{1}{2}\int d^{5}x\mathcal{N}^{Mb}_{a}\partial_{M}\psi^{a}_{b}\partial_{M}\psi^{a}_{b}.
\end{equation}
which upon variation gives the equation of motion for the shear mode gravitational fluctuations $\psi^{a}_{b}$
\begin{equation}\label{isoeom}
 \partial_{M}(\mathcal{N}^{Mb}_{a}(u)\partial_{M}\psi^{a}_{b})=0.
 \end{equation}
\subsection{Holographic RG flow equation for the shear viscosity tensor in isotropic bulk spacetime}
The fact that classical equations of motion in the bulk corresponds to RG flow equations in the field theory side was anticipated at the early stage of AdS/CFT \cite{Akhmedov:1998vf}. Therefore, for example, the holographic RG flow equations for the shear viscosity tensor $\eta^{b}\,_{a}\equiv\eta^{b}\,_{a}\,^{b}\,_{a}$, and conductivity $\sigma$ were derived, using the equations of motion for the scalar modes of the gravitational fluctuations, and Maxwell's equations of motion, respectively, for an electrically neutral isotropic black hole background \cite{Iqbal:2008by}, which were trivial in the hydrodynamic limit. And, recently, \cite{Sin:2011yh} has derived the same flow equation, for the conductivity, using the holographic Wilsonian renormalization group method \cite{Faulkner:2010jy,Heemskerk:2010hk}, and has provided the proof for the equivalence of the two methods in a general black hole background. Also, \cite{Matsuo:2011fk} has derived the holographic RG flow equation for $\sigma$, using the equations of motion for U(1) gauge fields in a charged black hole background, which is non-trivial even in the hydrodynamics limit, and is in agreement with the one derived in \cite{Chakrabarti:2010xy} using Kubo's formula.

Now, we derive the RG flow equation for the shear viscosity tensor $\eta^{b}\,_{a}\equiv\eta^{b}\,_{a}\,^{b}\,_{a}$, which is extracted from the correlation function $\langle T^{b}\,_{a}T^{b}\,_{a}\rangle$ where $T^{b}\,_{a}$ is dual to $h^{a}\,_{b}$, in isotropic bulk spacetime using the equation of motion (\ref{isoeom}). To this end, integrating by parts the bulk action (\ref{isosaction}), and using the equation of motion (\ref{isoeom}), we'll be left with the on-shell boundary action
 \begin{equation}\label{baction}
S_{eff}=-S_{B}[\epsilon],
\end{equation}
where the boundary action at $u=\epsilon$, $S_{B}[\epsilon]$, is
\begin{equation}\label{baction}
S_{B}[\epsilon]=-\frac{1}{2}\int_{u=\epsilon}d^{4}x\mathcal{N}^{ub}_{a}\psi^{a}_{b}\partial_{u}\psi^{a}_{b}.
\end{equation}
And, the canonical conjugate momentum along the radial direction $\Pi$ is
 \begin{equation}\label{isopi}
 \Pi=\frac{\delta S_{B}}{\delta\psi^{a}_{b}}=-\mathcal{N}^{ub}_{a}\partial_{u}\psi^{a}_{b}.
 \end{equation}
In terms of $\Pi$ (\ref{isopi}) the equation of motion (\ref{isoeom}) can be re-written, in the momentum space, as
 \begin{equation}\label{isopieom}
 \partial_{u}\Pi=-\bigl(\mathcal{N}^{tb}_{a}\omega^{2}+\mathcal{N}^{cb}_{a}k_{c}^{2}\bigr)\psi^{a}_{b}.
 \end{equation}
Note that $a\neq c$. The shear viscosity tensor $\eta^{b}\,_{a}$ is defined by $\eta^{b}\,_{a}\equiv\frac{\Pi}{i\omega\psi^{a}_{b}}$, and taking its first derivative with respect to $\epsilon$, we'll get
\begin{equation}\label{isopdeta}
\partial_{\epsilon}\eta^{b}\,_{a}=\frac{\partial_{u}\Pi}{i\omega\psi^{a}_{b}}-\frac{\Pi\partial_{u}\psi^{a}_{b}}{i\omega(\psi^{a}_{b})^{2}}.
\end{equation}
Then, using (\ref{isopieom}) and (\ref{isopi}) in (\ref{isopdeta}), we'll find the holographic RG flow equation for $\eta^{b}\,_{a}$ to be
\begin{equation}\label{isoflowetaz}
\partial_{\epsilon}\eta^{b}\,_{a}=i\omega\bigl(\frac{
(\eta^{b}\,_{a})^{2}}{\mathcal{N}^{ub}_{a}}+\mathcal{N}^{tb}_{a}\bigr)+\frac{i}{\omega}\mathcal{N}^{cb}_{a}k_{c}^{2},
\end{equation}
One can see that the RG flow equation (\ref{isoflowetaz}) is trivial in the hydrodynamics limit $k_{c}=0$, and $\omega\rightarrow0$. Hence, the shear viscosity tensor $\eta^{b}\,_{a}$ takes the same value at any hypersurface $u=\epsilon$. And, the initial data at the horizon is provided by requiring regularity at the horizon $\epsilon=u_{h}$ \cite{Iqbal:2008by}. Since $\frac{1}{\mathcal{N}^{ub}_{a}}$ and $\mathcal{N}^{tb}_{a}$ diverge at the horizon $\epsilon=u_{h}$, for the solution to be regular at the horizon, the right hand side of (\ref{isoflowetaz}) has to vanish at $\epsilon=u_{h}$. From which we recover, the frequency and momentum independent result
\begin{eqnarray}\label{isoetahax}
\eta^{b}\,_{a}(\epsilon=u_{h})&=&\sqrt{-\mathcal{N}^{ub}_{a}\mathcal{N}^{tb}_{a}}=\frac{1}{2\kappa^{2}}\sqrt{\frac{g(u_{h})}{g_{uu}(u_{h})g_{tt}(u_{h})}}\frac{g_{aa}(u_{h})}{g_{bb}(u_{h})}.
\end{eqnarray}
And, using the entropy density $s=\frac{1}{4G}\sqrt{\frac{g(u_{h})}{g_{uu}(u_{h})g_{tt}(u_{h})}}$, the shear viscosity to entropy density ratio at the horizon $\epsilon=u_{h}$ will be
\begin{equation}\label{isoetashab}
\frac{\eta^{b}\,_{a}(\epsilon=u_{h})}{s}=\frac{1}{4\pi}\frac{g_{aa}(u_{h})}{g_{bb}(u_{h})}=\frac{1}{4\pi},
\end{equation}
where we used the fact that $g_{aa}=g_{bb}$, for any $a$ and $b$ in isotropic spacetime. And, since the RG flow is trivial, in the hydrodynamic limit, the shear viscosity to entropy density ratio $\frac{\eta^{b}\,_{a}(\epsilon)}{s}$ will be given by (\ref{isoetashab}) at any hypersurface $u=\epsilon$, i.e.,
\begin{equation}\label{isoetashax2}
\frac{\eta^{b}\,_{a}(\epsilon)}{s}=\frac{\eta^{b}\,_{a}(\epsilon=u_{h})}{s}=\frac{\eta^{b}\,_{a}(\epsilon=0)}{s}=\frac{1}{4\pi}.
\end{equation}
This proves the universality of $\frac{\eta^{b}\,_{a}(\epsilon)}{s}$,
in isotropic bulk spacetime. But, we'll see, later on, that the universality of $\frac{\eta^{b}\,_{a}}{s}$ is no more valid in anisotropic bulk spacetime where different components of the shear viscosity tensor $\eta^{b}\,_{a}$, hence $\frac{\eta^{b}\,_{a}}{s}$, will take different values, and some components of it will RG flow non-trivially, i.e., their value at the horizon (IR) will be different from the one at the boundary (UV).

\section{\label{sec:aniso}Holographic RG flow in anisotropic bulk spacetime}
In this section, we study the holographic RG flow of the shear viscosity tensor in anisotropic $\mathcal{N}=4$ $SU(N_{c})$ super-Yang-Mills plasma by using its type IIB supergravity dual in anisotropic bulk spacetime derived in \cite{Mateos:2011tv}. The anisotropic version of an $\mathcal{N}=4$ $SU(N_{c})$ super-Yang-Mills plasma is given by deforming the gauge theory by the Chern-Simons term \cite{Azeyanagi:2009pr,Mateos:2011tv}
\begin{equation}
\delta S=\frac{1}{8\pi^2}\int\theta(z)\,\textnormal{Tr }F \wedge F,
\end{equation}
with $\theta (z)=2\pi az$ depending linearly on one of the spatial dimensions. The constant $\textit{a}$ is related to the density of D7-branes which are homogeneously distributed along $\textit{z}$ and are dissolved in the bulk of the dual theory \cite{Mateos:2011tv}.

\subsection{Effective action for the gravitational shear mode fluctuations in anisotropic bulk spacetime}
\label{sec:derivation}

Our five dimensional axion-dilaton gravity bulk action, which is a type IIB supergravity action where the Ramond-Ramond (RR) field, the axion, is a 0-form potential which is the 'magnetic' dual of the 8-form potential which couples to D7-branes 'electrically', is \cite{Azeyanagi:2009pr,Mateos:2011tv,Rebhan:2011vd}
\begin{equation}\label{bulkaction}
S_{\rm bulk}=\frac{1}{2\kappa^2}\int_{\mathcal{M}}\!\!\sqrt{-g}\,\Bigl(R+12-\frac{(\partial \phi)^2}{2}-e^{2\phi}\frac{(\partial \chi)^2}{2}\Bigr),
\end{equation}
where $\kappa^2=8\pi G=\frac{4\pi^2}{N_{c}^{2}}$.
The background solutions for the equation of motions resulting from the variation of this action are \cite{Mateos:2011tv}
\begin{equation}
\chi = az,
\end{equation}
\begin{eqnarray}\label{background}
ds^{2}&=&g_{MN}dx^{M}dx^{N}=g_{tt}dt^{2}+g_{aa}dx^{a}dx^{a}+g_{uu}du^{2}=g_{tt}dt^{2}+g_{ii}dx^{i}dx^{i}+g_{zz}dz^{2}+g_{uu}du^{2}\nonumber\\
&=&\frac{e^{-\phi(u)/2}}{u^2}\Big(-\mathcal{F}(u)\mathcal{B}(u)\,dt^2+\frac{du^2}{\mathcal{F}(u)}+dx^2+dy^2+\mathcal{H}(u)\,dz^2\Big)
\end{eqnarray}
Indices: $\{L, M, N,\}$ run over the full 5-dimensional bulk; $\{a,b,c\}$ run over all spatial coordinates $x$, $y$, and $z$; $\{i,j\}$ stand for $x$ and $y$ only. Also, throughout this paper the Einstein summation convention will apply only for indices $\{L, M, N,\}$ but not for $\{a,b,c\}$ and $\{i,j\}$. And,
\begin{eqnarray}
\phi (u) &=& -\frac{a^{2}u_h^{2}}{4}\log(1+\frac{u^{2}}{u_h^2})+O(a^4),\\
 \mathcal{F}(u)&=&1-\frac{u^4}{u_h^4}+\frac{a^2}{24u_h^{2}}[8u^{2}(u_h^{2}-u^{2})-10u^{4}\log2+(3u_h^{4}+7u^{4})\log(1+\frac{u^2}{u_h^{2}})]+O(a^4),\\
 \mathcal{B}(u)&=&1-\frac{a^{2}u_h^{2}}{24}[\frac{10u^{2}}{u_h^{2}+u^{2}}+\log(1+\frac{u^2}{u_h^{2}})]+O(a^4),\\
 \mathcal{H}(u) &=& e^{-\phi (u)},
\end{eqnarray}
for $a\ll T$. And, the horizon $u_h$ and the entropy density $s$ are related to the temperature $\textit{T}$ by \cite{Mateos:2011tv}
\begin{equation}\label{horizon}
u_{h}=\frac{1}{\pi T}+\frac{5\log2-2}{48\pi^{3}T^{3}}a^{2}+O(a^{4}),
\end{equation}
\begin{equation}\label{entropy}
s=\frac{1}{4G}\sqrt{\frac{g(u_{h})}{g_{uu}(u_{h})g_{tt}(u_{h})}}=\frac{\pi^{2}N_{c}^{2}T^{3}}{2}+\frac{N_{c}^{2}T}{16}a^{2}+O(a^{4}).
\end{equation}

Turning on only the metric fluctuations $h_{MN}$ about the background solution $g_{MN}^{0}$ (\ref{background}), i.e. $g_{MN}=g^{0}_{MN}+h_{MN}$, expanding the bulk action (\ref{bulkaction}) to second order in $h_{MN}$, and also using the gauge $h_{Mu}=0$, we'll have \cite{Rebhan:2011vd}
\begin{eqnarray}
S^{(2)}&=&\frac{1}{16\pi G}\int d^5 x \Bigl[\sqrt{-g}^{(2)}2A^{(0)}+\sqrt{-g}^{(0)}\left(
R^{(2)}-\frac{1}{2}e^{2\phi}a^2g^{zz(2)}\right)\Bigr],
\end{eqnarray}
where
\begin{eqnarray}
  A^{(0)} &=& -\frac{1}{2}(8+\frac{1}{2}\phi'^{2}g^{uu}+\frac{1}{2}e^{2\phi}a^{2}g^{zz})^{(0)},  \\
  g^{zz(2)} &=& g^{LL(0)}g^{zz(0)}g^{zz(0)}h_{Lz}h_{Lz}.
\end{eqnarray}

Using the trick of \cite{Kovtun:2003wp,Iqbal:2008by} of Kaluza-Klein dimensional reduction in the $a$ direction, considering only $h_{Na}=h_{Na}(x_{M}\neq a)$, and using the gauge $h_{NN}=h_{uN}=0$, we'll get the effective action
\begin{eqnarray}\label{action}
  S^{(2)}_{eff} &=& \int d^{5}x\bigl(-\frac{1}{4}\mathcal{N}^{MN}_{a}F^{a}_{MN}F^{a}_{MN}-\frac{1}{2}\mathcal{M}^{L}A^{z}_{L}A^{z}_{L}\bigr),
\end{eqnarray}
where
\begin{eqnarray}
F^{a}_{MN} &=& \partial_{M}A^{a}_{N}-\partial_{N}A^{a}_{M}, \\
 A^{a}_{N} &=& h^{a}\,_{N}=g^{aa(0)}h_{aN}, \\
 \mathcal{N}^{MN}_{a}(u) &=& \frac{1}{2\kappa^{2}}g^{(0)}_{aa}\sqrt{-g}^{(0)}g^{MM(0)}g^{NN(0)}, \\
 \mathcal{M}^{L}(u) &=& \frac{1}{2\kappa^2}a^{2}e^{2\phi}\sqrt{-g}^{(0)}g^{LL(0)}.
\end{eqnarray}
Note that this action, as emphasized in \cite{Iqbal:2008by}, is exactly in the form of the standard Maxwell's action with an effective coupling for the gauge fields
\begin{equation}\label{effectivecoupling}
  \frac{1}{g^{2}_{eff_a}}=\frac{1}{2\kappa^{2}}g^{(0)}_{aa}.
\end{equation}
It's obvious from the above relationship that the effective coupling $g_{eff_i}\neq g_{eff_z}$ since $g_{ii}\neq g_{zz}$. Hence, we have two distinct effective theories depending on which coupling and gauge fields we use. The gauge fields $A_{N}^{i}$ are coupled by $g_{eff_i}$, and the gauge fields $A_{N}^{z}$ are coupled by $g_{eff_z}$. For example, using the effective theory with the $g_{eff_i}$ we can extract the shear viscosity tensor $\eta^{b}\,_{i}\,^{b}\,_{i}$ from the correlation function $\langle T^{b}\,_{i}T^{b}\,_{i}\rangle$ where $T^{b}\,_{i}$ is dual to $h^{i}\,_{b}$. Similarly, using the effective theory with the $g_{eff_z}$ we can extract the shear viscosity tensor $\eta^{b}\,_{z}\,^{b}\,_{z}$ from the correlation function $\langle T^{b}\,_{z}T^{b}\,_{z}\rangle$ where $T^{b}\,_{z}$ is dual to $h^{z}\,_{b}$. Therefore, there are three independent components of the shear viscosity tensor $\eta^{b}\,_{a}\,^{b}\,_{a}$ , in the bulk, namely
\begin{eqnarray}
  \eta^{j}\,_{i}\equiv\eta^{j}\,_{i}\,^{j}\,_{i}&=&\eta^{x}\,_{y}\,^{x}\,_{y}=\eta^{y}\,_{x}\,^{y}\,_{x}, \nonumber\\
  \eta^{z}\,_{i}\equiv\eta^{z}\,_{i}\,^{z}\,_{i}&=&\eta^{z}\,_{x}\,^{z}\,_{x}=\eta^{z}\,_{y}\,^{z}\,_{y}, \nonumber\\
  \eta^{i}\,_{z}\equiv\eta^{i}\,_{z}\,^{i}\,_{z}&=&\eta^{x}\,_{z}\,^{x}\,_{z}=\eta^{y}\,_{z}\,^{y}\,_{z}.
\end{eqnarray}
However, we observe that two of the three independent components of the shear viscosity tensor in the bulk, $\eta^{z}\,_{i}$, and $\eta^{i}\,_{z}$, take the same value at the boundary, hence, we have only two independent components of the shear viscosity tensor at the boundary. This is consistent with the fact that the one index up and one index down energy-momentum tensor operator at the boundary is symmetric, and the shear viscosity tensor has only two independent components at the boundary \cite{Erdmenger:2010xm}.

Now, we start studying the properties of the shear viscosities using their corresponding effective actions. The effective action for $A^{z}_{i}$ with the effective gauge coupling $g_{eff_z}$ can be found from the action (\ref{action}) by setting $a=z$, $N=i$, and $L=i$
\begin{equation}\label{eaction1}
S^{(2)}_{eff}=\int d^{5}x\bigl(-\frac{1}{2}\mathcal{N}^{Mi}_{z}\partial_{M}\psi^{z}_{i}\partial_{M}\psi^{z}_{i}-\frac{1}{2}\mathcal{M}^{i}\psi^{z}_{i}\psi^{z}_{i}\bigr)
\end{equation}
where
\begin{eqnarray}
\mathcal{N}^{Mi}_{z}&=&\frac{1}{2\kappa^{2}}g^{(0)}_{zz}\sqrt{-g}^{(0)}g^{MM(0)}g^{ii(0)},\\
\mathcal{M}^{i} &=& \frac{1}{2\kappa^2}a^{2}e^{2\phi}\sqrt{-g}^{(0)}g^{ii(0)},\\
\psi^{z}_{i}(t,u,y)&=&A^{z}_{i}(t,u,y)=h^{z}\,_{i}(t,u,y).
\end{eqnarray}

Similarly, the effective action for $A^{i}_{b}$ with the effective gauge coupling $g_{eff_i}$ can be found from the action (\ref{action}) by setting $a=i$, $N=b$, and $L=b$
\begin{equation}\label{eaction3}
S^{(2)}_{eff}=\int d^{5}x\bigl(-\frac{1}{2}\mathcal{N}^{Mb}_{i}\partial_{M}\psi^{i}_{b}\partial_{M}\psi^{i}_{b}\bigr)
\end{equation}
where
\begin{eqnarray}
\mathcal{N}^{Mb}_{i}&=&\frac{1}{2\kappa^{2}}g^{(0)}_{ii}\sqrt{-g}^{(0)}g^{MM(0)}g^{bb(0)},\\
\mathcal{M}^{i} &=& \frac{1}{2\kappa^2}a^{2}e^{2\phi}\sqrt{-g}^{(0)}g^{ii(0)},\\
\psi^{i}_{b}(t,u,z)&=&A_{b}^{i}(t,u,z)=h^{i}\,_{b}(t,u,z).
\end{eqnarray}
Note that we have dropped the mass-like term $\frac{1}{2}\mathcal{M}^{b}\psi^{z}_{b}\psi^{z}_{b}$ from (\ref{eaction3}) since it doesn't affect the equation of motion for $\psi^{i}_{b}$. Also, since (\ref{eaction3}) is the same effective action as the isotropic one (\ref{isoaction}) discussed in the previous section, we can immediately observe that $\eta^{b}\,_{i}$ has a trivial RG flow, and the components of $\frac{\eta^{b}\,_{i}}{s}$ take the values
\begin{equation}\label{universal}
\frac{\eta^{j}\,_{i}(\epsilon)}{s}=\frac{1}{4\pi}\frac{g_{ii}}{g_{jj}}=\frac{1}{4\pi},
\end{equation}
and,
\begin{equation}\label{etash2}
\frac{\eta^{z}\,_{i}(\epsilon)}{s}=\frac{1}{4\pi}\frac{g_{ii}(u_{h})}{g_{zz}(u_{h})}=\frac{1}{4\pi\mathcal{H}(u_h)}=\frac{1}{4\pi}\bigl(1-\frac{\log2}{4\pi^{2}}\bigl(\frac{a}{T}\bigr)^{2}+O(a^{4})\bigr)<\frac{1}{4\pi}
\end{equation}
for $a\neq0$. Equations (\ref{universal}), and (\ref{etash2}) are exactly Eq.14, and Eq.17 of reference \cite{Rebhan:2011vd}, respectively, derived using the membrane paradigm approach.

But, in order to calculate $\eta^{i}\,_{z}$ one has to solve the RG flow equation that we'll get from the corresponding effective action (\ref{eaction1}).

\subsection{Holographic RG flow equation for the shear viscosity tensor in anisotropic bulk spacetime}\label{sec:asymptotic}

In this section, using the equation of motion for the shear modes of gravitational fluctuations, we derive the holographic RG flow equation for the shear viscosity $\eta^{i}\,_{z}$. Varying the effective action (\ref{eaction1}), we find the equation of motion
 \begin{equation}\label{eom}
 \partial_{M}(\mathcal{N}^{Mi}_{z}\partial_{M}\psi^{z}_{i})-\mathcal{M}^{i}\psi^{z}_{i}=0.
 \end{equation}
Using the equation of motion (\ref{eom}) in the bulk action (\ref{eaction1}), we get the on-shell action
\begin{equation}\label{baction}
S^{(2)}_{eff}=-S_{B}[\epsilon],
\end{equation}
where the boundary action at $u=\epsilon$, $S_{B}[\epsilon]$, is
\begin{equation}\label{baction}
S_{B}[\epsilon]=-\frac{1}{2}\int_{u=\epsilon}d^{4}x\mathcal{N}^{ui}_{z}\psi^{z}_{i}\partial_{u}\psi^{z}_{i}.
\end{equation}
And, the canonical conjugate momentum along the radial direction $\Pi$ is
 \begin{equation}\label{pi}
 \Pi=\frac{\delta S_{B}}{\delta\psi^{z}_{i}}=-\mathcal{N}^{ui}_{z}\partial_{u}\psi^{z}_{i}.
 \end{equation}
In terms of $\Pi$ (\ref{pi}) the equation of motion (\ref{eom}) can be re-written, in the momentum space, as
 \begin{equation}\label{pieom}
 \partial_{u}\Pi=-\bigl(\mathcal{N}^{ti}_{z}\omega^{2}+\mathcal{N}^{yi}_{z}k_{y}^{2}+\mathcal{M}^{i}\bigr)\psi^{z}_{i}.
 \end{equation}
The shear viscosity tensor $\eta^{i}\,_{z}$ is defined by $\eta^{i}\,_{z}\equiv\frac{\Pi}{i\omega\psi^{z}_{i}}$, and taking its first derivative with respect to $\epsilon$, we'll get
\begin{equation}\label{pdeta}
\partial_{\epsilon}\eta^{i}\,_{z}=\frac{\partial_{u}\Pi}{i\omega\psi^{z}_{i}}-\frac{\Pi\partial_{u}\psi^{z}_{i}}{i\omega(\psi^{z}_{i})^{2}}.
\end{equation}
Then, using (\ref{pieom}) and (\ref{pi}) in (\ref{pdeta}), we find the holographic RG flow equation for $\eta^{i}\,_{z}$ to be
\begin{equation}\label{flowetazj}
\partial_{\epsilon}\eta^{i}\,_{z}=i\omega\bigl(\frac{
(\eta^{i}\,_{z})^{2}}{\mathcal{N}^{ui}_{z}}+\mathcal{N}^{ti}_{z}\bigr)+\frac{i}{\omega}\bigl(\mathcal{N}^{yi}_{z}k_{y}^{2}+\mathcal{M}^{i}\bigr),
\end{equation}
which is non trivial even in the hydrodynamics limit $k_{y}=0$ and $\omega\rightarrow0$. One can also see that at $a=0$, which makes $\mathcal{M}^{i}=0$, the flow equation (\ref{flowetazj}) reduces to the isotropic one (\ref{isoflowetaz}). The flow equation (\ref{flowetazj}) can also be derived by using the holographic Wilsonian RG method, and Kubo's formula as shown in Appendix \ref{sec:wilson}, and Appendix \ref{sec:kubo}, respectively.

\section{\label{sec:soln}Solution}

\begin{figure}
 \begin{center}
\includegraphics[width=0.60\textwidth]{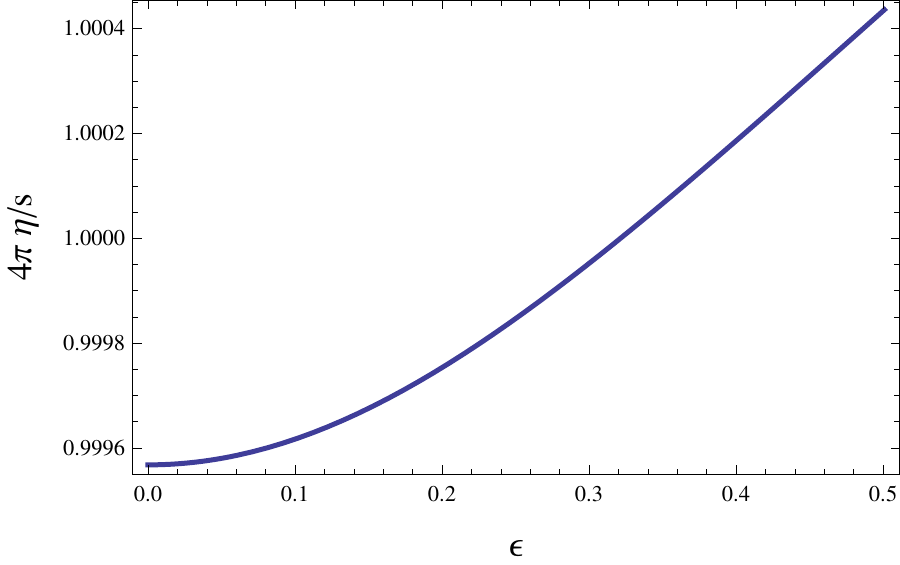}
\caption{Shear viscosity $\eta\equiv\Re(\eta^{i}\,_{z})$ over $s/4\pi$ as a function of the radial coordinate $\epsilon$ with $u_{h}=0.50$, $a=0.1$, and $T=0.64$.
\label{fig:etase222222}}
 \end{center}
\end{figure}

In this section, we solve the flow equations (\ref{flowetazj}) analytically up to second order in the anisotropy parameter $a$. The initial data at the horizon is provided by requiring regularity at the horizon $\epsilon=u_{h}$ \cite{Iqbal:2008by}. Since $\frac{1}{\mathcal{N}^{ui}_{z}}$ and $\mathcal{N}^{ti}_{z}$ diverge at $\epsilon=u_{h}$, in order for the solution to be regular at the horizon, the right hand side of (\ref{flowetazj}) has to vanish at $\epsilon=u_{h}$. From which we recover frequency, momentum and mass-like term $\mathcal{M}^{i}$ independent result
\begin{eqnarray}\label{etaha1}
\eta^{i}\,_{z}(\epsilon=u_{h})&=&\sqrt{-\mathcal{N}^{ui}_{z}\mathcal{N}^{ti}_{z}}=\frac{1}{2\kappa^{2}}\sqrt{\frac{g(u_{h})}{g_{uu}(u_{h})g_{tt}(u_{h})}}\frac{g_{zz}(u_{h})}{g_{ii}(u_{h})}.
\end{eqnarray}
And, using (\ref{entropy}), the shear viscosity to entropy density ratio at the horizon $\epsilon=u_{h}$ will be
\begin{equation}\label{etashaa1}
\frac{\eta^{i}\,_{z}(\epsilon=u_{h})}{s}=\frac{1}{4\pi}\frac{g_{zz}(u_{h})}{g_{ii}(u_{h})}=\frac{1}{4\pi}\mathcal{H}(u_h)=\frac{1}{4\pi}\bigl(1+\frac{\log2}{4\pi^{2}}\bigl(\frac{a}{T}\bigr)^{2}+O(a^{4})\bigr)>\frac{1}{4\pi}
\end{equation}
for $a\neq0$. Writing out $\eta^{j}\,_{z}=\Re(\eta^{i}\,_{z})+i\Im(\eta^{i}\,_{z})$ in (\ref{flowetazj}), taking $\omega\rightarrow 0$ limit, setting $k_{y}=0$, and writing out the metric components explicitly, we'll get
\begin{eqnarray}
\partial_{\epsilon}\Im(\eta^{i}\,_{z})-\frac{a^{2}}{2\kappa^{2}\omega}\frac{e^{\frac{3}{4}\phi(\epsilon)}\sqrt{\mathcal{B(\epsilon)}}}{\epsilon^{3}} &=& 0, \\
 \label{reeta}\partial_{\epsilon}\Re(\eta^{i}\,_{z})+4\omega\kappa^{2}\frac{e^{\frac{9}{4}\phi(\epsilon)}\epsilon^{3}}{\mathcal{F(\epsilon)}\sqrt{\mathcal{B(\epsilon)}}}\Im(\eta^{i}\,_{z})\Re(\eta^{i}\,_{z}) &=& 0.
\end{eqnarray}
Since, we are interested only up to second order in $a$, we'll take $\mathcal{B}=e^{\phi}=1+O(a^2)$, and $\mathcal{F}(u)=1-\frac{u^4}{u_h^4}+O(a^2)=\frac{(u_{h}^{2}+u^{2})(u_{h}^{2}-u^{2})}{u_h^4}+O(a^2)$. Therefore, up to a second order in $a$, the flow equation for $\Im(\eta^{i}\,_{z})$ can be written as
\begin{eqnarray}\label{imeta}
\partial_{\epsilon}\Im(\eta^{i}\,_{z})&=&\frac{a^{2}}{2\kappa^{2}\omega}\frac{1}{\epsilon^{3}}+O(a^{4}).
\end{eqnarray}
Solving (\ref{imeta}), using the initial condition at the horizon $\Im(\eta^{j}\,_{z}(\epsilon=u_{h}))=0$, and using it in (\ref{reeta}), we'll get
\begin{equation}\label{reeta2}
 \partial_{\epsilon}\Re(\eta^{i}\,_{z})-\bigl[\frac{u_{h}^{2}\epsilon}{\epsilon^{2}+u_{h}^{2}}a^{2}+O(a^{4})\bigr]\Re(\eta^{i}\,_{z})=0.
\end{equation}
Note that $\omega$ is canceled out. Solving (\ref{reeta2}), and setting $\Re(\eta^{i}\,_{z})\equiv\eta(\epsilon)$, we'll get
\begin{eqnarray}\label{original}
\eta(\epsilon)&=&\eta(u_h)\bigl(\frac{\epsilon^{2}+u_{h}^2}{2u_{h}^2}\bigr)^{\frac{a^2u_h^2}{2}}+O(a^4)=\eta(u_h)\bigl(1+\frac{1}{2}a^2u_h^2\log[\frac{\epsilon^{2}+u_{h}^2}{2u_{h}^2}]\bigr)+O(a^4),
\end{eqnarray}
which, after using (\ref{etaha1}), and (\ref{horizon}), becomes
\begin{equation}\label{eta}
\eta(\epsilon)=\frac{\pi N_{c}^{2}T^{3}}{8}+\bigl(1+\log[\frac{1}{4}(1+\pi^{2}T^{2}\epsilon^{2})^{4}]\bigr)\frac{N_{c}^{2}T}{64\pi}a^{2}+O(a^{4}).
\end{equation}
Note that at $a=0$ (\ref{eta}) reduces to the isotropic case calculated in \cite{Policastro:2001yc}. And, using (\ref{entropy}), the shear viscosity to entropy density ratio at any hypersurface $u=\epsilon$ will be
\begin{equation}\label{mainflow}
\frac{\eta(\epsilon)}{s}=\frac{1}{4\pi}\bigl(1+\frac{\log[\frac{1}{2}(1+\pi^{2}T^{2}\epsilon^{2})^{2}]}{4\pi^{2}}\bigl(\frac{a}{T}\bigr)^{2}+O(a^{4})\bigr).
\end{equation}
Note again that when $a=0$ in (\ref{mainflow}) $\frac{\eta(\epsilon)}{s}$ will take the universal value $\frac{1}{4\pi}$. We've plotted the holographic RG flow of $\frac{\eta(\epsilon)}{s}$ (\ref{mainflow}), for a fixed value of $a$ and $T$, in Fig.~\ref{fig:etase222222}.
\begin{figure}
 \begin{center}
\includegraphics[width=0.60\textwidth]{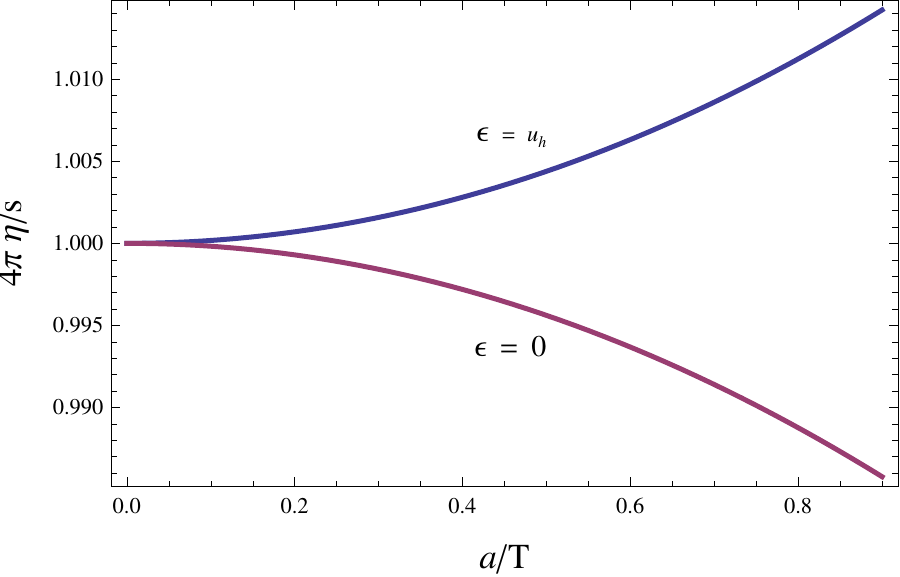}
\caption{Shear viscosity $\eta\equiv\Re(\eta^{i}\,_{z})$ over $s/4\pi$ as a function of the anisotropy parameter
$a/T$ at the horizon $\epsilon=u_{h}=0.50$, and at the boundary $\epsilon=0$ for $a\ll T$.
\label{fig:etastnew}}
 \end{center}
\end{figure}

 As we can see from (\ref{mainflow}), the shear viscosity to entropy density ratio at the boundary $\epsilon=0$ becomes
\begin{equation}\label{etasb}
\frac{\eta(\epsilon=0)}{s}=\frac{1}{4\pi}\bigl(1-\frac{\log2}{4\pi^{2}}\bigl(\frac{a}{T}\bigr)^{2}+O(a^{4})\bigr)<\frac{1}{4\pi}.
\end{equation}
Note that (\ref{etasb}) is equivalent to (\ref{etash2}) as advertised in section \ref{sec:derivation}. And, at the horizon $\epsilon^{2}=u_{h}^{2}=\frac{1}{\pi^{2}T^{2}}$, (\ref{mainflow}) reproduces (\ref{etashaa1}), as expected,
\begin{eqnarray}\label{etasha}
\frac{\eta(\epsilon=u_{h})}{s}&=&\frac{1}{4\pi}\bigl(1+\frac{\log2}{4\pi^{2}}\bigl(\frac{a}{T}\bigr)^{2}+O(a^{4})\bigr)>\frac{1}{4\pi}.
\end{eqnarray}
We've plotted the temperature flows of $\frac{\eta(\epsilon=u_{h})}{s}$ (\ref{etasha}), and $\frac{\eta(\epsilon=0)}{s}$ (\ref{etasb}) in Fig.~\ref{fig:etastnew}.

\section{\label{sec:conclusion}Conclusion}

We have revisited the calculation of the shear viscosities of the anisotropic, strongly coupled $\mathcal{N}=4$ super-Yang-Mills plasma by means of its type IIB supergravity dual which was previously carried out in \cite{Rebhan:2011vd} using membrane paradigm and numerical methods. We have showed that, at finite UV cut-off, there is an additional shear viscosity in addition to the other shear viscosities studied in \cite{Rebhan:2011vd}. Unlike, the shear viscosities studied in \cite{Rebhan:2011vd}, we have showed that our additional shear viscosity has a non-trivial RG flow equation. We have derived and solved the RG equation, analytically up to second order in the anisotropy parameter $a$, and have found that its value at the boundary (UV) is equivalent to one of the shear viscosities studied in \cite{Rebhan:2011vd}, and violates the holographic shear viscosity (Kovtun-Son-Starinets) bound.

Finally, we emphasize that, our observation, at the bulk away from the boundary, the shear viscosity tensor has three independent components due to the antisymmetry of the one index up and one index down energy-momentum tensor operator, while at the boundary, it only has two, since the one index up and one index down energy-momentum tensor at the boundary is symmetric, is a theoretically interesting finding that calls for a better understanding of why at finite UV cut-off the structure of the theory changes qualitatively.

\acknowledgments
The author thanks A. Lewis Licht, Bo Ling, and Misha Stephanov for reading the draft.

\appendix

\section{\label{sec:wilson} Derivation of the holographic RG flow equation for the shear viscosity $\eta^{i}\,_{z}$ using the holographic Wilsonian RG method}
In this section, we derive the holographic RG flow equation of $\eta^{i}\,_{z}$ (\ref{flowetazj}) using the holographic Wilsonian RG method following closely reference \cite{Faulkner:2010jy}, and \cite{Sin:2011yh}. Holographic Wilsonian RG formulation for the gravity dual of strongly coupled isotropic $\mathcal{N}=4$ super-Yang-Mills plasma was developed in \cite{Faulkner:2010jy,Heemskerk:2010hk}, and more recently in \cite{Grozdanov:2011aa} where they mapped the problem of integrating out the boundary degrees of freedom above a cut-off scale $\Lambda$ to integrating out the bulk degrees of freedom below the cut-off hypersurface at $u = \epsilon$. And, integrating out the bulk degrees of freedom in the the region $u<\epsilon$ resulted in a boundary action $S_{B}(u=\epsilon)$ at $u=\epsilon$ hypersurface. They also proposed that $S_{B}$ can be identified with the Wilsonian effective action of the boundary theory at the scale $\Lambda$, with couplings in $S_{B}$ identified with those for single-trace and multi-trace operators in the boundary theory. Requiring that physical observable be independent of the choice of the cut-off scale $\epsilon$ then determined a flow equation for the Wilsonian action $S_{B}$ and associated couplings. The flow equation for those couplings was also given in \cite{Akhmedov:2002gq}.

In the following, we apply the formalism of \cite{Faulkner:2010jy} to the gravitational fluctuations in an anisotropic background. Our effective action (\ref{eaction1}), after integrating out the bulk modes below $u<\epsilon$ becomes
\begin{equation}
S=\int_{u>\epsilon} d^{5}x\mathcal{L}^{(2)}_{eff}(\psi,\partial_{M}\psi)+S_{B}[\psi,\epsilon]
\end{equation}
where
\begin{eqnarray}
\mathcal{L}^{(2)}_{eff}(\psi,\partial_{M}\psi)&=&-\frac{1}{2}\mathcal{N}^{Mx}_{z}\partial_{M}\psi\partial_{M}\psi-\frac{1}{2}\mathcal{M}^{x}\psi^{2},\\
\mathcal{N}^{Mx}_{z}&=&\frac{1}{2\kappa^{2}}\sqrt{-g}^{(0)}g^{(0)}_{zz}g^{MM(0)}g^{xx(0)},\\
\mathcal{M}^{x} &=& \frac{1}{2\kappa^2}a^{2}e^{2\phi}\sqrt{-g}^{(0)}g^{xx(0)},\\
\psi(t,u,y)=\psi^{z}_{x}&=&A_{x}^{z}(t,u,y)=A_{j}^{z}(t,u,c).
\end{eqnarray}
 Varying the action, we'll find the equation of motion
 \begin{equation}\label{weom}
 \partial_{M}(\mathcal{N}^{Mx}_{z}\partial_{M}\psi)-\mathcal{M}^{x}\psi=0
 \end{equation}
 with boundary condition at $u=\epsilon$
 \begin{equation}\label{pi2}
 \Pi=\frac{\delta S_{B}}{\delta \psi}, \qquad \Pi=-\mathcal{N}^{ux}_{z}\partial_{u}\psi
 \end{equation}
 where $\Pi$ is the canonical momentum along the radial direction.

 The flow equation for $S_{B}[\psi,\epsilon]$ can be determined by requiring $\frac{d}{d\epsilon}S=0$ \cite{Faulkner:2010jy}:
 \begin{equation}\label{flowsb1}
 \partial_{\epsilon}S_{B}[\psi,\epsilon]=-\int_{u=\epsilon} d^{4}x\bigl(\Pi\partial_{\epsilon}\psi-\mathcal{L}^{(2)}_{eff}\bigr)=-\int_{u=\epsilon} d^{4}x\mathcal{H},
 \end{equation}
where $\mathcal{H}$ is the hamiltonian density for evolution in the $u$ direction. Writing out $\mathcal{H}$ explicitly and using (\ref{pi2}) we can write the flow equation as
\begin{eqnarray}
\partial_{\epsilon}S_{B}[\psi,\epsilon]&=&-\int_{u=\epsilon} d^{4}x\bigl(-\frac{1}{2\mathcal{N}^{ux}_{z}}(\frac{\delta S_{B}}{\delta \psi})^{2}+\frac{1}{2}\mathcal{N}^{\mu x}_{z}\partial_{\mu}\psi \partial_{\mu}\psi+\frac{1}{2}\mathcal{M}^{x}\psi^{2}\bigr),
\end{eqnarray}
where $\mu$, $\nu$ $\equiv$ $t$, $x$, $y$, $z$. Expanding $S_{B}$ in momentum space as \cite{Faulkner:2010jy,Sin:2011yh}
\begin{equation}\label{sb}
S_{B}[\psi,\epsilon]=-\frac{1}{2}\int\frac{d^{d}k}{(2\pi)^{d}}G(k,\epsilon)\psi(k)\psi(-k),
\end{equation}
where $G(k,\epsilon)$ is the Green's function at the cut-off hypersurface $u=\epsilon$. Also, note that we are considering only single trace deformation, and has set the couplings to the double trace operators to zero. Inserting (\ref{sb}) back to the flow equation (\ref{flowsb1}), and comparing the coefficients, one can obtain the flow equation for the Green's function $G(k,\epsilon)$
 \begin{equation}\label{flowg1}
 \partial_{\epsilon}G=-\frac{G^{2}}{\mathcal{N}^{ux}_{z}}+\mathcal{N}^{tx}_{z}\omega^{2}+\mathcal{N}^{yx}_{z}k_{y}^{2}+\mathcal{M}^{x}.
 \end{equation}
Defining the shear viscosity $\eta^{x}\,_{z}$ as \cite{Iqbal:2008by}
\begin{equation}
\eta^{x}\,_{z}\equiv\frac{\Pi}{-\partial_{t}\psi}=\frac{\Pi}{i\omega\psi}=-\frac{G}{i\omega},
\end{equation}
where we used $\Pi=\frac{\delta S_{B}}{\delta \psi}=-G\psi$ to get the last line. And, using it in (\ref{flowg1}), we'll get the flow equation for $\eta^{x}\,_{z}$
\begin{equation}\label{wflowetaz}
\partial_{\epsilon}\eta^{x}\,_{z}=i\omega\bigl(\frac{
(\eta^{x}\,_{z})^{2}}{\mathcal{N}^{ux}_{z}}+\mathcal{N}^{tx}_{z}\bigr)+\frac{i}{\omega}\bigl(\mathcal{N}^{yx}_{z}k_{y}^{2}+\mathcal{M}^{x}\bigr),
\end{equation}
which is exactly the same as equation (\ref{flowetazj}) with $i=x$.

\section{\label{sec:kubo} Derivation of the holographic RG flow equation for the shear viscosity $\eta^{i}\,_{z}$ using Kubo's formula}
 An equation of motion similar to (\ref{eom}) has appeared, in \cite{Chakrabarti:2010xy,Jain:2010ip}, as an equation of motion for U(1) gauge field $\phi=A_{x}$ in a charged black hole background. And, at zero frequency and zero momentum limit, it is given by Eq. 2.5 of \cite{Chakrabarti:2010xy}, i.e.,
  \begin{equation}\label{keom}
 \partial_{r}(N(r)\partial_{r}\phi)+M(r)\phi=0.
\end{equation}
where $N(r)$, and $M(r)$ are some functions of the radial direction $r$. Then, starting from Kubo's formula, as shown rigorously in reference \cite{Jain:2010ip}, one can find the real part of the conductivity $\Re(\sigma(r))$, at any radial distance $r$, in terms of the solution of (\ref{keom}), to be
\begin{equation}\label{cond}
 \Re(\sigma(r))=\sigma_{H}(\frac{\phi(r=r_{H})}{\phi(r)})^{2},
\end{equation}
where $\sigma_{H}$ is the value of the conductivity at the horizon $r=r_{H}$. Using this result, reference \cite{Chakrabarti:2010xy} has derived the RG flow equation for the conductivity $\sigma(r)$, which is given in Eq. A.15 of \cite{Chakrabarti:2010xy}, i.e.,
\begin{equation}\label{condflow}
\partial_{r}\sigma=i\omega\bigl(\frac{
2\kappa^{2}\sigma^{2}}{N}+\frac{1}{2\kappa^{2}}Ng_{rr}g^{tt}\bigr)-\frac{1}{2\kappa^{2}}\frac{i}{\omega}M
\end{equation}

Therefore, if we just replace $N$ by $2\kappa^{2}\mathcal{N}^{uj}_{z}$, $Ng_{rr}g^{tt}$ by $2\kappa^{2}\mathcal{N}^{ti}_{z}$, $M$ by $-2\kappa^{2}\mathcal{M}^{i}$, and $\sigma$ by $\eta^{i}\,_{z}$ in (\ref{condflow}), we will get (\ref{flowetazj}) with $k_{y}=0$. One should also note that, at zero momentum and zero frequency limit, (\ref{keom}) is exactly the same as (\ref{eom}) with these replacements.

\bibliographystyle{JHEP}
\bibliography{ref}

\begin{thebibliography}{10}

\bibitem{Maldacena:1997re}
  J.~M.~Maldacena,
  Adv.\ Theor.\ Math.\ Phys.\  {\bf 2}, 231 (1998)
  [hep-th/9711200].

\bibitem{Gubser:1998bc}
  S.~S.~Gubser, I.~R.~Klebanov and A.~M.~Polyakov,
  Phys.\ Lett.\ B {\bf 428}, 105 (1998)
  [hep-th/9802109].

\bibitem{Witten:1998qj}
  E.~Witten,
  Adv.\ Theor.\ Math.\ Phys.\  {\bf 2}, 253 (1998)
  [hep-th/9802150].


\bibitem{Susskind:1998dq}
  L.~Susskind and E.~Witten,
  hep-th/9805114.
  A.~W.~Peet and J.~Polchinski,
  Phys.\ Rev.\ D {\bf 59}, 065011 (1999)
  [hep-th/9809022].

\bibitem{de Haro:2000xn}
  S.~de Haro, S.~N.~Solodukhin and K.~Skenderis,
  Commun.\ Math.\ Phys.\  {\bf 217}, 595 (2001)
  [hep-th/0002230].
  K.~Skenderis,
  Class.\ Quant.\ Grav.\  {\bf 19}, 5849 (2002)
  [hep-th/0209067].
  M.~Bianchi, D.~Z.~Freedman and K.~Skenderis,
  Nucl.\ Phys.\ B {\bf 631}, 159 (2002)
  [hep-th/0112119].
  K.~Skenderis and B.~C.~van Rees,
  JHEP {\bf 0905}, 085 (2009)
  [arXiv:0812.2909 [hep-th]].

\bibitem{Son:2002sd}
  D.~T.~Son and A.~O.~Starinets,
  JHEP {\bf 0209}, 042 (2002)
  [hep-th/0205051].
  C.~P.~Herzog and D.~T.~Son,
  JHEP {\bf 0303}, 046 (2003)
  [hep-th/0212072].

\bibitem{Adams:2005dq}
  J.~Adams {\it et al.}  [STAR Collaboration],
  Nucl.\ Phys.\ A {\bf 757}, 102 (2005);
  [nucl-ex/0501009].
  K.~Adcox {\it et al.}  [PHENIX Collaboration],
  Nucl.\ Phys.\ A {\bf 757}, 184 (2005).
  [nucl-ex/0410003].

\bibitem{Romatschke:2007mq}
  P.~Romatschke and U.~Romatschke,
  Phys.\ Rev.\ Lett.\  {\bf 99}, 172301 (2007);
  [arXiv:0706.1522 [nucl-th]].
  H.~Song, {\it et al.},
  Phys.\ Rev.\ Lett.\  {\bf 106}, 192301 (2011).
  [arXiv:1011.2783 [nucl-th]].

\bibitem{Kovtun:2004de}
  P.~Kovtun, D.~T.~Son and A.~O.~Starinets,
  Phys.\ Rev.\ Lett.\  {\bf 94}, 111601 (2005)
  [hep-th/0405231].

\bibitem{Damour:1979de}
  T. Damour,
  Th`ese de Doctorat d�Etat, Universit�e Pierre
  et Marie Curie, Paris VI, 1979.

\bibitem{Policastro:2001yc}
  G.~Policastro, D.~T.~Son and A.~O.~Starinets,
  Phys.\ Rev.\ Lett.\  {\bf 87}, 081601 (2001)
  [hep-th/0104066].

\bibitem{Kovtun:2003wp}
  P.~Kovtun, D.~T.~Son and A.~O.~Starinets,
  JHEP {\bf 0310}, 064 (2003)
  [hep-th/0309213].

\bibitem{Iqbal:2008by}
  N.~Iqbal and H.~Liu,
  Phys.\ Rev.\ D {\bf 79}, 025023 (2009)
  [arXiv:0809.3808 [hep-th]].

\bibitem{Kovtun:2011np}
  P.~Kovtun, G.~D.~Moore and P.~Romatschke,
  Phys.\ Rev.\ D {\bf 84}, 025006 (2011)
  [arXiv:1104.1586 [hep-ph]].

\bibitem{Myers:2009ij}
  R.~C.~Myers, M.~F.~Paulos and A.~Sinha,
  JHEP {\bf 0906}, 006 (2009);
  [arXiv:0903.2834 [hep-th]].
  S.~Cremonini and P.~Szepietowski,
  JHEP {\bf 1202}, 038 (2012);
  [arXiv:1111.5623 [hep-th]].
  A.~Buchel and S.~Cremonini,
  JHEP {\bf 1010}, 026 (2010).
  [arXiv:1007.2963 [hep-th]].

\bibitem{Cremonini:2011iq}
  S.~Cremonini,
  Mod.\ Phys.\ Lett.\ B {\bf 25}, 1867 (2011)
  [arXiv:1108.0677 [hep-th]].

\bibitem{Basu:2009vv}
  P.~Basu, {\it et al.},
  Phys.\ Lett.\ B {\bf 689}, 45 (2010);
  [arXiv:0911.4999 [hep-th]].
  M.~Ammon, {\it et al.},
  Phys.\ Lett.\ B {\bf 686}, 192 (2010).
  [arXiv:0912.3515 [hep-th]].
  M.~Natsuume and M.~Ohta,
  Prog.\ Theor.\ Phys.\  {\bf 124}, 931 (2010);
  [arXiv:1008.4142 [hep-th]].
  P.~Basu and J.~-H.~Oh,
  JHEP {\bf 1207}, 106 (2012)
  arXiv:1109.4592 [hep-th].

\bibitem{Erdmenger:2010xm}
  J.~Erdmenger, P.~Kerner and H.~Zeller,
  Phys.\ Lett.\ B {\bf 699}, 301 (2011);
  [arXiv:1011.5912 [hep-th]].

\bibitem{Florkowski:2008ag}
  W.~Florkowski,
  Phys.\ Lett.\ B {\bf 668}, 32 (2008)
  [arXiv:0806.2268 [nucl-th]].
  W.~Florkowski and R.~Ryblewski,
  Acta Phys.\ Polon.\ B {\bf 40}, 2843 (2009)
  [arXiv:0901.4653 [nucl-th]].

\bibitem{Azeyanagi:2009pr}
  T.~Azeyanagi, W.~Li and T.~Takayanagi,
  JHEP {\bf 0906}, 084 (2009)
  [arXiv:0905.0688 [hep-th]].

\bibitem{Mateos:2011tv}
  D.~Mateos and D.~Trancanelli,
  JHEP {\bf 1107}, 054 (2011)
  [arXiv:1106.1637 [hep-th]].

\bibitem{Rebhan:2011vd}
  A.~Rebhan and D.~Steineder,
  Phys.\ Rev.\ Lett.\  {\bf 108}, 021601 (2012)
  [arXiv:1110.6825 [hep-th]].

\bibitem{Faulkner:2010jy}
  T.~Faulkner, H.~Liu and M.~Rangamani,
  JHEP {\bf 1108}, 051 (2011)
  [arXiv:1010.4036 [hep-th]].
\bibitem{Heemskerk:2010hk}
  I.~Heemskerk and J.~Polchinski,
  JHEP {\bf 1106}, 031 (2011)
  [arXiv:1010.1264 [hep-th]].
\bibitem{Grozdanov:2011aa}
  S.~Grozdanov,
  JHEP {\bf 1206}, 079 (2012)
  [arXiv:1112.3356 [hep-th]].

\bibitem{Akhmedov:2002gq}
  E.~T.~Akhmedov,
  hep-th/0202055.

\bibitem{Akhmedov:1998vf}
  E.~T.~Akhmedov,
  Phys.\ Lett.\ B {\bf 442}, 152 (1998)
  [hep-th/9806217].

\bibitem{Sin:2011yh}
  S.~-J.~Sin and Y.~Zhou,
  JHEP {\bf 1105}, 030 (2011)
  [arXiv:1102.4477 [hep-th]].

\bibitem{Matsuo:2011fk}
  Y.~Matsuo, S.~-J.~Sin and Y.~Zhou,
  JHEP {\bf 1201}, 130 (2012)
  [arXiv:1109.2698 [hep-th]].

\bibitem{Jain:2010ip}
  S.~Jain,
JHEP {\bf 1011}, 092 (2010)
[arXiv:1008.2944 [hep-th]].

\bibitem{Chakrabarti:2010xy}
  S.~K.~Chakrabarti, S.~Chakrabortty and S.~Jain,
  JHEP {\bf 1102}, 073 (2011)
  [arXiv:1011.3499 [hep-th]].

\bibitem{Oh:2012zu}
  J.~-H.~Oh,
  JHEP {\bf 1206}, 103 (2012)
  [arXiv:1201.5605 [hep-th]].


\bibitem{Carroll:1990bd}
  S.~Carroll, G.~Field and R.~Jackiw,
  Phys.\ Rev.\ D {\bf 41}, 1231 (2009).


\bibitem{Landsteiner:2007bd}
  K.~Landsteiner and J.~Mas,
  JHEP {\bf 0707}, 088 (2007)
  [arXiv:0706.0411 [hep-th]].

\bibitem{Adams:2008wt}
  A.~Adams, K.~Balasubramanian and J.~McGreevy,
  JHEP {\bf 0811}, 059 (2008)
  [arXiv:0807.1111 [hep-th]].



\end{thebibliography}

\end{document}